\documentclass[pre,showpacs,twocolumn,floatfix,bibnotes]{revtex4}
\usepackage{here}
\usepackage[dvips]{graphicx}

\newcommand\pictc[5]{\begin{figure}
                       \centerline{
                       \includegraphics[width=#1\columnwidth]{#3}}
                   \protect\caption{\protect\label{fig:#4} #5}
                    \end{figure}            }
\newcommand\pict[4][1.]{\pictc{#1}{!tb}{#2}{#3}{#4}}
\newcommand\rpict[1]{\ref{fig:#1}}
\newcommand\leqt[1]{\protect\label{eq:#1}}
\newcommand\reqtn[1]{\ref{eq:#1}}
\newcommand\reqt[1]{(\reqtn{#1})}

\newcounter{Fig}

\begin{document}

\begin{sloppy}

\title{Nonlinear properties of left-handed metamaterials}

\author{Alexander A. Zharov$^{1,2}$, Ilya V. Shadrivov$^1$, and Yuri S. Kivshar$^1$}

\affiliation{$^1$Nonlinear Physics Group, Research School of
Physical Sciences and Engineering, Australian National
University, Canberra ACT 0200, Australia \\
$^2$ Institute for Physics of Microstructures, Russian Academy of
Sciences, Nizhny Novgorod 603950, Russia}

\begin{abstract}
We analyze nonlinear properties of microstructured materials with
negative refraction, the so-called {\em left-handed
metamaterials}. We consider a two-dimensional periodic structure
created by arrays of wires and split-ring resonators embedded into
a nonlinear dielectric, and calculate the effective nonlinear
electric permittivity and magnetic permeability. We demonstrate
that the hysteresis-type dependence of the magnetic permeability
on the field intensity allows changing the material from left- to right-handed and back. These effects can be treated as {\em the second-order phase transitions} in the transmission properties induced by the variation of an external field.
\end{abstract}


\maketitle

Recent theoretical studies \cite{Pendry:1996-4773:PRL,
Pendry:1999-2075:ITMT,markos} and experimental results
\cite{Smith:2000-4184:PRL,bay,prl_new} have shown the possibility
to create novel types of microstructured materials which
demonstrate the property of negative refraction. In particular,
the composite materials created by arrays of wires and split-ring
resonators were shown to possess the negative real part of the
magnetic permeability and dielectric permittivity for microwaves.
These materials are often referred to as {\em left-handed
materials} or {\em materials with negative refraction}. Properties
of such materials were analyzed theoretically by Veselago long
time ago \cite{veselago}, but only recently they were demonstrated
experimentally. As was shown by Veselago \cite{veselago}, the
left-handed materials possess a number of peculiar properties,
including negative refraction for the interface scattering,
inverse light pressure, reverse Doppler and Vavilov-Cherenkov
effects.

So far, all properties of left-handed materials were studied in
the linear regime of wave propagation when both magnetic
permeability and dielectric permittivity of the material are
assumed to be independent on the intensity of the electromagnetic
field. However, any future effort in creating {\em tunable
structures} where the field intensity changes the transmission
properties of the composite structure would require the knowledge
of nonlinear properties of such metamaterials which may be quite
unusual. In this Letter we analyze, for the first time to our
knowledge, {\em nonlinear properties of left-handed metamaterials}
for the example of a lattice of the split-ring resonators and wires with a nonlinear dielectric. We show that the effective magnetic permeability depends on the intensity of the macroscopic magnetic field in a nontrivial way, allowing
{\em switching between the left- and right-handed materials by
varying the field intensity}. We believe that our findings may
stimulate the future experiments in this field, as well as the
studies of nonlinear effects in photonic crystals where the
phenomenon of negative refraction is analyzed now very intensively
\cite{pbg,pbg1}.

We consider a two-dimensional composite structure consisting of a
square lattice of the periodic arrays of conducting wires and
split-ring resonators (SRR), shown schematically in
Fig.~\rpict{geom}. We assume that the unit-cell size $d$ of the
structure is much smaller then the wavelength of the propagating
electromagnetic field and, for simplicity,  we choose the
single-ring geometry of a lattice of cylindrical SRRs. The results obtained for
this case are qualitatively similar to more involved cases of
double SRRs properties. This type of the microstructured materials has
recently been suggested and built in order to create left-handed
metamaterials with negative refraction in the microwave
region~\cite{Smith:2000-4184:PRL}.

Negative real part of the effective dielectric permittivity of
such a composite structure appears due to the metallic wires
whereas a negative sign of the magnetic permeability becomes
possible due to the SRR lattice. As a result, these materials
demonstrate the properties of negative refraction in the finite
frequency band, $\omega_{0} < \omega < {\rm min} (\omega_p,
\omega_{\parallel m})$, where $\omega_{0}$ is the
eigenfrequency of SRRs, $\omega_{\parallel m}$ is the frequency of
the longitudinal magnetic plasmon, $\omega_p$ is an effective
plasma frequency, and $\omega$ is an angular frequency of the
propagating electromagnetic wave, $({\cal E},{\cal H})\sim ({\bf
E},{\bf H})\exp{(i\omega t)}$. The split-ring resonator can be
described as an effective LC oscillator (see Ref.
\cite{Gorkunov:2002-263:EPB}) with capacitance of the SRR gap, as well as effective inductance, and resistance (see the upper insert in Fig.~\rpict{geom}).

{\em Nonlinear response} of such a composite structure can be
characterized by two different contributions. The first one is an
intensity-dependent part of the effective dielectric permittivity
of the infilling dielectric. For simplicity, we assume that the
metallic structure is embedded into a nonlinear dielectric with
permittivity that depends on the intensity of electric field in a
general form,  $ \epsilon_D = \epsilon_D(|{\bf E}|^2)$. For
detailed calculations presented below, we take the linear
dependence that corresponds to the Kerr nonlinearity.

The second contribution into the nonlinear properties of the
composite material comes from the lattice of resonators, since the
SRR capacitance (and, therefore, the SRR eigenfrequency) depends
on the strength of a local electric field in a narrow slot.
Additionally, we can expect a nonlinear eigenfrequency detuning
due to a resonant growth of the charge density at the edges of the
SRR gap. The intensity of the local electric field in the SRR gap
depends on the electromotive force in the resonator loop, which is
induced by the magnetic field. Therefore, the effective magnetic
permeability $\mu_{\rm eff}$ should depend on the macroscopic
(average) magnetic field ${\bf H}$.

\pict{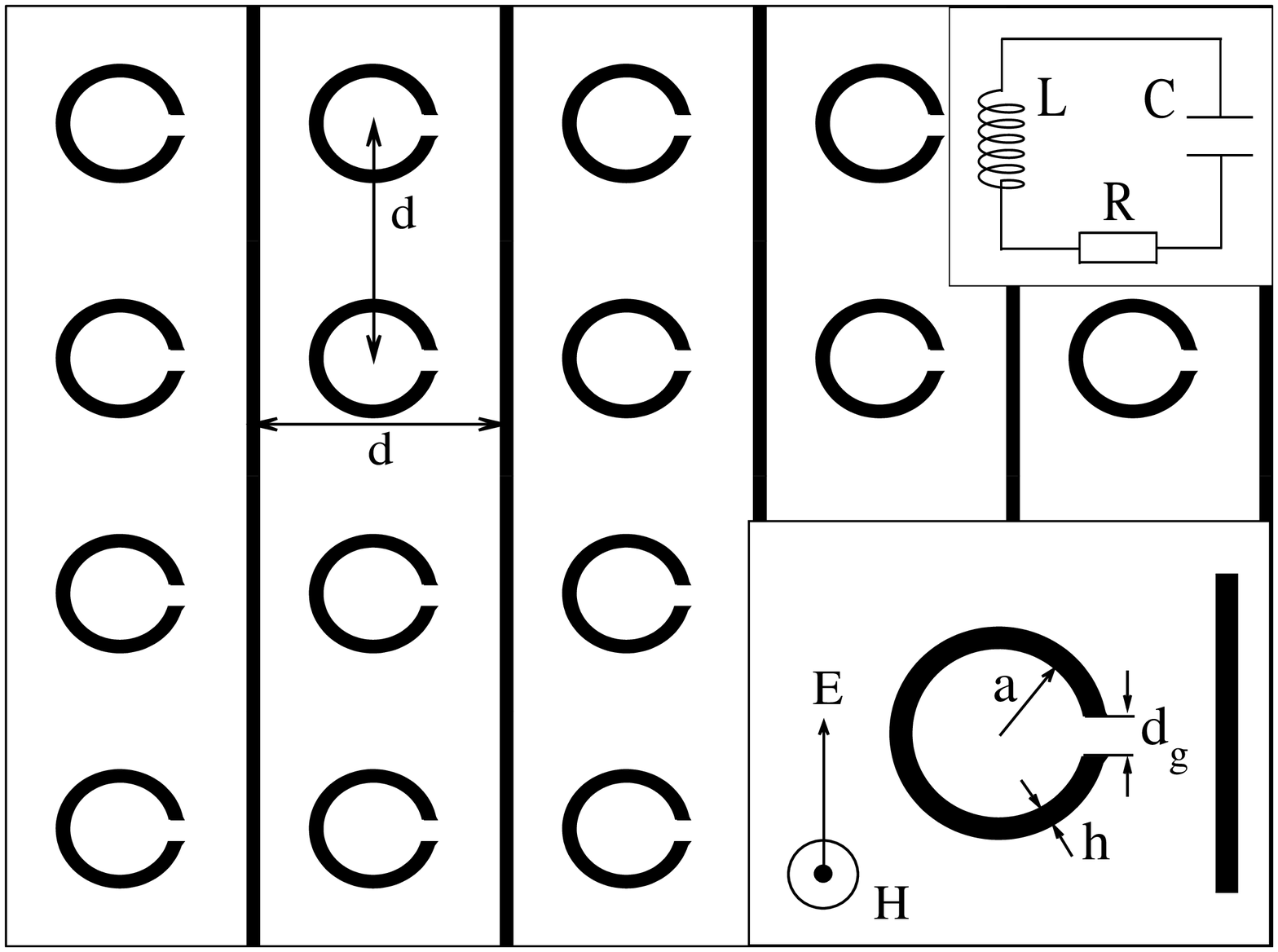}{geom}{Schematic of the composite metamaterial
structure. Lower insert shows a unit cell of the periodic
structure while the upper inset shows the SRR equivalent
oscillator with the parameters used in the derivation.}

For the polarization shown in Fig.~\rpict{geom} (lower insert),
the main contribution to the dielectric function is given by the
array of wires. When the wire length is large enough, so that the
frequency of the fundamental (dipole) mode of an individual wire
becomes much less than $\omega$, only the resistance and inductance of the wires give a contribution into the lattice impedance. Hence, the Ohm law for
the current can be written in the form,
\begin{equation} \leqt{ohm}
{\bf j}_{\omega} \approx \frac{\sigma}{1+i\omega \sigma S L_w}{\bf
E}^{\prime},
\end{equation}
where ${\bf j}_{\omega}$ is the electric current density in the
wire, ${\bf E}^{\prime}$ is the local electric field, $\sigma$ is
the conductivity of the wire metal, $L_w \approx 2c^{-2}
\ln{(d/r)}$ ($d \gg r$) is the inductance of the wire per unit
length, $c$ is the speed of light, $r$ is the wire radius, $S$ is the effective area of the wire cross-section, $S \approx \pi r^2$, for $\delta>r$ and $S
\approx \pi\delta(2r-\delta)$, for $\delta<r$, where $\delta = c/ \sqrt{( 2\pi\sigma\omega) }$ is the thickness of the skin layer. The average current density in the unit sell can be written in the form,
\begin{equation} \leqt{cur_density}
\langle {\bf j}_{\omega} \rangle = \frac{S}{d^2}{\bf j}_{\omega}.
\end{equation}

For the waves polarized along the wires the average macroscopic
electric field ${\bf E}$ is approximately equal to the local field
${\bf E}^{\prime}$. Taking into account the general relation
between the electric field ${\bf E}$ and the electric induction
${\bf D}$,
\begin{equation} \leqt{induction}
{\bf D} = \epsilon_D(|{\bf E}|^2) {\bf E} + \frac{4\pi}{i\omega}
\langle {\bf j}_{\omega} \rangle,
\end{equation}
we can obtain the expression for the  effective nonlinear
dielectric permittivity
\begin{equation}
\leqt{permittivity}
\epsilon_{\rm eff} \left( |E|^2 \right) = \epsilon_{D}\left( |E|^2
\right) - \frac{\omega_p^2}{\omega(\omega-i\gamma_{\epsilon})},
\end{equation}
where $\omega_p \approx (c/d) [2\pi/\ln{(d/r)}]^{1/2}$ is the
effective plasma frequency, and $\gamma_{\epsilon} = c^2/2\sigma S
\ln{(d/r)}$. The second term in the right-hand side of Eq.
\reqt{permittivity} is in a full agreement with the earlier result
obtained by Pendry and co-authors~\cite{Pendry:1996-4773:PRL}. One should note
that the low losses case, i.e. $\gamma_{\epsilon} \ll \omega$,
corresponds to the condition $\delta \ll r$.

The analysis becomes more involved for calculating {\em a
nonlinear magnetic response} of the composite structure, which is
determined by the intrinsic properties of  the interacting
nonlinear oscillators in the presence of an external periodic
force. For the structure under consideration, the current induced
in each resonator can be found as
\begin{equation} \leqt{current}
I=- i \pi a^2 \left(\frac{\omega}{c}\right) \left| {\bf H}_0
\right| Z^{-1} ,
\end{equation}
where ${\bf H}_0$ is the amplitude of the external field applied
to SRR, $Z = i\omega L + R +(i\omega C)^{-1}$ is the SRR
impedance, and other parameters are
marked in Fig.~\rpict{geom}: $a$ is the radius, $L$, $R$, and $C$
are  inductance, resistance and capacitance, respectively.  The
amplitude of the electric field in the gap of SRR can be found
approximately as follows
\begin{equation} \leqt{loc_field}
\left | {\bf E}_g \right| \approx \frac{ I }{ i \omega C d_g },
\end{equation}
where $d_g$ is the size of the SRR gap. Nonlinear effects in
Eqs.~\reqt{current},~\reqt{loc_field} appear due to the
capacitance $C$ which is proportional to $\epsilon_{D}(|E_g|^2)$.
Therefore, Eq.~\reqt{loc_field} gives an implicit relation between
the amplitude of the local electric field in the gap and the
amplitude of the (external to SRR) magnetic field.

The relation between the local (microscopic) and average
(macroscopic) magnetic fields can be obtained by averaging a local
field over the volume of the unit cell. 
\begin{equation} \leqt{integr_3}
\langle {\bf B} \rangle = \langle {\bf H} \rangle + F {\bf H}^{\prime},
\end{equation}
where ${\bf H}^{\prime}$ is an additional magnetic field induced
by the alternating external magnetic field ${\bf H}_0$ in
cylindrical SRR,  which determines magnetization of the composite,
and $F= \pi a^2/d^2$.

The average macroscopic field can be also determined as follows,
\begin{equation} \leqt{H_average}
{\bf H} \equiv \langle {\bf H} \rangle = (1-F){\bf H}_0 + F({\bf
H}_0+{\bf H}^{\prime}) = {\bf H}_0 + F{\bf H}^{\prime}.
\end{equation}
Taking into account that ${\bf H}^{\prime}=0$ outside SRR, from
the boundary conditions we obtain the relation
\begin{equation} \leqt{bound_cond}
\left|{\bf H}^{\prime} \right| = \frac{4\pi}{c}\left|{\bf j}_s \right|,
\end{equation}
where ${\bf j}_s$ is an equivalent surface current in SRR, which
is equal to the current per unit length. From Eq. \reqt{current}
and Eqs. \reqt{integr_3}-\reqt{bound_cond}, we obtain an explicit
expression for the effective magnetic permeability of the
composite structure (for $F \ll 1$):
\begin{equation} \leqt{mu_eff}
\mu_{\rm eff}({\bf H}) = 1 + \frac{F\,
\omega^2}{\omega_{0NL}^2({\bf H}) -
                \omega^2 + i \Gamma \omega},
\end{equation}
where
\[
\omega_{0NL}^2({\bf H})= \left(\frac{c}{a}\right)^2
\frac{d_g}{[\pi (1-F) h \epsilon_D(|{\bf E}_g({\bf H})|^2)]}
\]
is the eigenfrequency of oscillations in the presence of the
external field of a finite amplitude, $h$ is the width of the
ring, $\Gamma=c^2/2\pi(1-F)\sigma ah$,  for $h<\delta$,  and
$\Gamma=c^2/2\pi(1-F)\sigma a\delta$, for $h>\delta$. It is
important to note that Eq.~\reqt{mu_eff} has a simple physical
interpretation: The resonant frequency of the artificial magnetic
structure depends on the amplitude of the external magnetic field
and, in turn, this leads to the intensity-dependent function
$\mu_{\rm eff}$.

\pict{fig02.eps}{Re_mu}{Real part of the effective magnetic
permeability vs.  intensity of the magnetic field: (a)~$\Omega>1$,
$\alpha=1$; (b)~$\Omega<1$, $\alpha=1$, (c)~$\Omega>1$,
$\alpha=-1$; and (d)~$\Omega<1$, $\alpha=-1$. Black -- the
lossless case ($\gamma = 0$), grey--the lossy case ($\gamma =
0.05$). Dashed curves show unstable branches.}
\pict{fig03.eps}{Im_mu}{Imaginary part of the effective magnetic
permeability vs.  intensity of the magnetic field for $\gamma =
0.05$: (a)~$\Omega>1$, $\alpha=1$; (b)~$\Omega<1$, $\alpha=1$,
(c)~$\Omega>1$, $\alpha=-1$; and (d)~$\Omega<1$, $\alpha=-1$.
Dashed curves show unstable branches.}

To be more specific, we consider the Kerr nonlinearity of the
dielectric in the composite material, i.e
\begin{equation} \leqt{kerr_epsilon}
\epsilon_{D}\left( |E|^2 \right) = \epsilon_{D0} + \alpha |E|^2/E_c^2,
\end{equation}
where $E_c$ is a characteristic electric field, $\alpha=\pm 1$
stands for the  focusing or defocussing nonlinearity, respectively.
Then, the relation between the macroscopic magnetic field and the
dimensionless nonlinear resonant frequency can be obtained from
Eqs.~\reqt{current}--\reqt{bound_cond}, and \reqt{kerr_epsilon} as
\begin{equation} \leqt{magn_field}
|{\bf H}|^2 = \alpha A^2E_c^2 \frac{ \left( 1-X^2 \right)\left[
                                \left(X^2-\Omega^2\right)
                                +\Omega^2\gamma^2 \right]}{X^6},
\end{equation}
where $A^2=16\epsilon_{D0}^3(1-F)\omega_{0}^2h^2/c^2$,
$\Omega=\omega/\omega_{0}$, $\omega_{0}=(c/a)[d_g / \pi (1-F) h
\epsilon_{D0}]^{1/2}$ is the eigenfrequency of the system of SRRs
in the linear limit, $X = \omega_{0NL}/\omega_{0}$, and $\gamma =
\Gamma/\omega_{0}$. Therefore, we find that the dimensionless
eigenfrequency of the SRR lattice $X\left(|H|^2\right)$ is a
multi-valued function of the magnetic field. This result reveals a
general property of nonlinear oscillators with a high quality
factor \cite{rabinovich}.

Parametric dependence of the effective magnetic permeability on
the magnetic field is determined completely by Eqs.~\reqt{magn_field}
and  \reqt{mu_eff}. Figures ~\rpict{Re_mu} and \rpict{Im_mu}
summarize different types of nonlinear magnetic properties of the
composite, which are defined by the dimensionless frequency of the
external field $\Omega$,  for both {\em focusing} [Fig.
\rpict{Re_mu}(a,b) and Fig. \rpict{Im_mu}(a,b)] and {\em
defocussing} [Fig.~\rpict{Re_mu}(c,d) and Fig. \rpict{Im_mu} (c,d)]
nonlinearity of the dielectric.

In the case of {\em focusing nonlinearity} (i.e. when $\alpha
=1$), the SRR eigenfrequency decreases with the intensity of the
electromagnetic field because of a growth of the SRR capacitance.
Then, for $\Omega>1$, the effective magnetic permeability of the
composite structure grows with the field intensity, as shown in
Fig.~\rpict{Re_mu}(a). If in the linear limit the composite
material is left-handed, i.e. $Re(\mu_{\rm eff})<0$, it will
become right-handed for higher intensities of the magnetic field
[in this reasoning we assume $Re(\epsilon_{\rm eff})<0$].

More complicated behavior of the magnetic permeability is observed
for $\Omega<1$; this is shown in Fig.~\rpict{Re_mu}(b). Here, in
the linear limit the real part of the magnetic permeability is
always positive, but the eigenfrequency of SRR decreases with the
growth of the magnetic field, thus driving the system into the
resonance. Since a nonlinear oscillator has a hysteresis structure
of its response with a change of an external force (see, e.g.,
Ref. \cite{rabinovich}), this leads to multi-valued dependencies.
In our problem,  the nonlinear eigenfrequency is a three-valued
function of the external magnetic field, and this results in jumps
of the magnetic permeability with the growth of the magnetic
field. As follows from Fig.~\rpict{Re_mu}(b), the magnetic field
intensity displays a jump of the magnetic permeability from
positive to negative values at some $H_{c1}$. Thus, the initially
opaque medium with positive refraction becomes a
negative-refraction transparent medium with the growth of the
field intensity. This effect can be treated as the second-order
phase transition induced by the external electromagnetic field.
Reversed transition takes place when the magnetic field intensity
decreases to the value $H_{c2}<H_{c1}$.

In the case of {\em defocussing nonlinearity} (i.e. when
$\alpha=-1$), the SRR eigenfrequency increases with the amplitude
of the external field. That is why the resonance effects take
place for $\Omega>1$, as shown in Figs. \rpict{Re_mu}(c,d) and
Figs. \rpict{Im_mu}(c,d). Here, we observe the opposite behavior
when the transition from the case $Re(\mu_{\rm eff})<0$ to the
case $Re(\mu_{\rm eff})>0$ takes place at high values of the
external field, and the reversed transition occurs at lower field
intensities. In the latter case, $Re(\mu_{\rm eff})$ is always
positive for $\Omega<1$, see Fig.~\rpict{Re_mu}(d).

Our results show that the imaginary part of the effective magnetic
permeability, which determines the structure losses, can be
controlled rather effectively by a proper choice of the intensity
of the external high-frequency magnetic field, see
Fig.~\rpict{Im_mu}. We believe that this feature may be important
for the future applications of left-handed materials.

Due to the high values of the electric field in the gap of SRR and
the resonant interaction of the electromagnetic field with the SRR
lattice, the characteristic magnetic nonlinearity in such
structures is much stronger then the corresponding electric
nonlinearity. Therefore, {\em the magnetic nonlinearity should
dominate in the composite materials which display the phenomenon
of negative refraction}. Moreover, nonlinear elements can be used only for filling the gap in SRR, allowing an easy tuning by applying an external field.

The possibility of strongly enhanced nonlinearities in left-handed metamaterials revealed here may lead to an essential revision of the concepts based on the linear theory, since the electromagnetic waves propagating in such
materials always have a finite amplitude.  At the same time, the
engineering of highly nonlinear composite materials will open a
number of novel opportunities for their microwave applications,
such as frequency multipliers, beam spatial spectrum transformers,
switchers, limiters, etc.

It is important to note that the hysteresis behavior with jumps in the dependencies of the effective material parameters has been described above for the stationary processes only.  Such transitions will display a
characteristic scale in time or space for initial or boundary
problems, respectively. Such spatial or temporal scales are
determined by the relaxation micro-processes in the SRR lattice.

In conclusion, we have presented, for the first time to our
knowledge, a systematic analysis of nonlinear properties of the
microstructured materials which display negative refraction, the
so-called left-handed metamaterials.  We have shown that the
composite metamaterials composed of a lattice of wires and
split-ring resonators possess an effective magnetic permeability
which depends on the intensity of the macroscopic magnetic field
in a nontrivial way. The magnetic nonlinearity is found to be much
stronger than the nonlinearity in the dielectric properties due to
the field enhancement in the split-ring resonators. The dependence
of the effective magnetic permeability on the field intensity
allows switching between its positive and negative values, i.e. a
change of the material properties from left- to right-handed and
back. Such processes can be treated as the second-order phase
transitions induced by the variation of an external
electromagnetic field.

We thanks Costas Soukoulis for useful discussions and suggestions.
This work was partially supported by the Australian Research
Council and the US Air Force-Far East Office.

\end{sloppy}
\end{document}